\documentclass[fleqn,usenatbib]{mnras}

\usepackage{newtxtext,newtxmath}

\usepackage[T1]{fontenc}

\DeclareRobustCommand{\VAN}[3]{#2}
\let\VANthebibliography\thebibliography
\def\thebibliography{\DeclareRobustCommand{\VAN}[3]{##3}\VANthebibliography}

\usepackage{graphicx}	
\usepackage{amsmath}	
\usepackage{ragged2e}

\title[A triply-imaged variable AGN at $z=2.06$]{A variable active galactic nucleus at $z=2.06$ triply-imaged by the galaxy cluster MACS~J0035.4-2015}

\author[L. J. Furtak et al.]{
Lukas J. Furtak,$^{1}$\thanks{E-mail:~furtak@post.bgu.ac.il}
Ramesh Mainali,$^{2,3,4}$
Adi Zitrin,$^{1}$
Ad\`{e}le Plat,$^{3}$
Seiji Fujimoto,$^{5}$
Megan Donahue,$^{6}$
\newauthor Erica J. Nelson,$^{7}$
Franz E. Bauer,$^{8,9,10}$
Ryosuke Uematsu,$^{11}$
Gabriel B. Caminha,$^{12}$
Felipe Andrade-Santos,$^{13,14}$
\newauthor Larry D. Bradley,$^{15}$
Karina I. Caputi,$^{16,17}$
St\'{e}phane Charlot,$^{18}$
Jacopo Chevallard,$^{19}$
Dan Coe,$^{15,20,21}$
\newauthor Emma Curtis-Lake,$^{22}$
Daniel Espada,$^{23,24}$
Brenda L. Frye,$^{3}$
Kirsten K. Knudsen,$^{25}$
Anton M. Koekemoer,$^{15}$
\newauthor Kotaro Kohno,$^{26,27}$
Vasily Kokorev,$^{16}$
Nicolas Laporte,$^{28,29}$
Minju M. Lee,$^{17,30}$
Brian C. Lemaux,$^{31,32}$
\newauthor Georgios E. Magdis,$^{17,30,33}$
Keren Sharon,$^{34}$
Daniel P. Stark,$^{3}$
Yuanyuan Su,$^{35}$
Katherine A. Suess,$^{36,37}$
\newauthor Yoshihiro Ueda,$^{11}$
Hideki Umehata,$^{38,39}$
Alba Vidal-Garc\'{i}a$^{40,41}$
and John F. Wu,$^{15,21}$
\\
$^{1}$Physics Department, Ben-Gurion University of the Negev, P.O. Box 653, Be'er-Sheva 84105, Israel\\
$^{2}$Observational Cosmology Lab, NASA Goddard Space Flight Center, Greenbelt, MD 20771, USA\\
$^{3}$Steward Observatory, University of Arizona, 933 N Cherry Ave, Tucson, AZ, USA\\
$^{4}$Department of Physics, The Catholic University of America, Washington, DC 20064, USA\\
$^{5}$Department of Astronomy, The University of Texas at Austin, Austin, TX 78712, USA\\
$^{6}$Michigan State University, Physics \& Astronomy Department, East Lansing, MI, USA\\
$^{7}$Department for Astrophysical and Planetary Science, University of Colorado, Boulder, CO 80309, USA\\
$^{8}$Instituto de Astrof{\'{\i}}sica, Facultad de F{\'{i}}sica, Pontificia Universidad Cat{\'{o}}lica de Chile, Campus San Joaquín, Av. Vicuña Mackenna 4860, Macul Santiago,\\
Chile, 7820436\\
$^{9}$Centro de Astroingenier{\'{\i}}a, Facultad de F{\'{i}}sica, Pontificia Universidad Cat{\'{o}}lica de Chile, Campus San Joaquín, Av. Vicuña Mackenna 4860, Macul Santiago,\\
Chile, 7820436\\
$^{10}$Millennium Institute of Astrophysics, Nuncio Monse{\~{n}}or S{\'{o}}tero Sanz 100, Of 104, Providencia, Santiago, Chile\\
$^{11}$Department of Astronomy, Kyoto University, Kyoto 606-8502, Japan\\
$^{12}$Max-Planck-Institut f\"{u}r Astrophysik, Karl-Schwarzschild-Str. 1, 85748 Garching, Germany\\
$^{13}$Department of Liberal Arts and Sciences, Berklee College of Music, 7 Haviland Street, Boston, MA 02215, USA\\
$^{14}$Center for Astrophysics | Harvard \& Smithsonian, 60 Garden Street, Cambridge, MA 02138, USA\\
$^{15}$Space Telescope Science Institute, 3700 San Martin Dr., Baltimore, MD 21218, USA\\
$^{16}$Kapteyn Astronomical Institute, University of Groningen, P.O. Box 800, 9700AV Groningen, The Netherlands\\
$^{17}$Cosmic Dawn Center (DAWN), Copenhagen, Denmark\\
$^{18}$Institut d'Astrophysique de Paris, CNRS, Sorbonne Universit\'e, 98bis Boulevard Arago, 75014, Paris, France\\
$^{19}$Department of Physics, University of Oxford, Denys Wilkinson Building, Keble Road, Oxford OX1 3RH, UK\\
$^{20}$Association of Universities for Research in Astronomy (AURA) for the European Space Agency (ESA), STScI, Baltimore, MD, USA\\
$^{21}$Center for Astrophysical Sciences, Department of Physics and Astronomy, The Johns Hopkins University, 3400 N Charles St. Baltimore, MD 21218, USA\\
$^{22}$Centre for Astrophysics Research, Department of Physics, Astronomy and Mathematics, University of Hertfordshire, Hatfield AL10 9AB, UK\\
$^{23}$Departamento de F\'{i}sica Te\'{o}rica y del Cosmos, Campus de Fuentenueva, Edificio Mecenas, Universidad de Granada, E-18071, Granada, Spain\\
$^{24}$Instituto Carlos I de F\'{i}sica Te\'{o}rica y Computacional, Facultad de Ciencias, E-18071, Granada, Spain\\
$^{25}$Chalmers University of Technology, Department of Space, Earth and Environment, SE-412 96 Gothenburg, Sweden\\
$^{26}$Institute of Astronomy, Graduate School of Science, The University of Tokyo, 2-21-1 Osawa, Mitaka, Tokyo 181-0015, Japan\\
$^{27}$Research Center for the Early Universe, Graduate School of Science, The University of Tokyo, 7-3-1 Hongo, Bunkyo-ku, Tokyo 113-0033, Japan\\
$^{28}$Kavli Institute for Cosmology, University of Cambridge, Madingley Road, Cambridge CB3 0HA, UK\\
$^{29}$Cavendish Laboratory, University of Cambridge, 19 JJ Thomson Avenue, Cambridge CB3 0HE, UK\\
$^{30}$DTU-Space, Technical University of Denmark, Elektrovej 327, DK2800 Kgs. Lyngby, Denmark\\
$^{31}$Department of Physics and Astronomy, University of California, Davis, One Shields Ave., Davis, CA 95616, USA\\
$^{32}$Gemini Observatory, NSF’s NOIRLab, 670 N. A’ohoku Place, Hilo, Hawai’i, 96720, USA\\
$^{33}$Niels Bohr Institute, University of Copenhagen, Jagtvej 128, 2100, Copenhagen N, Denmark\\
$^{34}$Department of Astronomy, University of Michigan, 1085 S. University Ave, Ann Arbor, MI 48109, USA\\
$^{35}$University of Kentucky, 505 Rose Street, Lexington, KY 40506, USA\\
$^{36}$Department of Astronomy and Astrophysics, University of California, Santa Cruz, 1156 High Street, Santa Cruz, CA 95064 USA\\
$^{37}$Kavli Institute for Particle Astrophysics and Cosmology and Department of Physics, Stanford University, Stanford, CA 94305, USA\\
$^{38}$Institute for Advanced Research, Nagoya University, Furocho, Chikusa, Nagoya 464-8602, Japan\\
$^{39}$Department of Physics, Graduate School of Science, Nagoya University, Furocho, Chikusa, Nagoya 464-8602, Japan\\
$^{40}$Observatorio Astron\'{o}mico Nacional, C/ Alfonso XII 3, 28014 Madrid, Spain\\
$^{41}$\'{E}cole Normale Sup\'{e}rieure, CNRS, UMR 8023, Universit\'{e} PSL, Sorbonne Universit\'{e}, Universit\'{e} de Paris, F-75005 Paris, France
}

\date{Accepted 2023 April 27. Received 2023 April 27; in original form 2023 March 2}

\pubyear{2023}

\begin{document}
\label{firstpage}
\pagerange{\pageref{firstpage}--\pageref{lastpage}}
\maketitle

\begin{abstract}
We report the discovery of a triply imaged active galactic nucleus (AGN), lensed by the galaxy cluster MACS~J0035.4-2015 ($z_{\mathrm{d}}=0.352$). The object is detected in \textit{Hubble Space Telescope} imaging taken for the RELICS program. It appears to have a quasi-stellar nucleus consistent with a point-source, with a de-magnified radius of $r_e\lesssim100$\,pc. The object is spectroscopically confirmed to be an AGN at $z_{\mathrm{spec}}=2.063\pm0.005$ showing broad rest-frame UV emission lines, and is detected in both X-ray observations with \textit{Chandra} and in ALCS ALMA~band 6 (1.2\,mm) imaging. It has a relatively faint rest-frame UV luminosity for a quasar-like object, $M_{\mathrm{UV},1450}=-19.7\pm0.2$. The object adds to just a few quasars or other X-ray sources known to be multiply lensed by a galaxy cluster. Some diffuse emission from the host galaxy is faintly seen around the nucleus and there is a faint object nearby sharing the same multiple-imaging symmetry and geometric redshift, possibly an interacting galaxy or a star-forming knot in the host. We present an accompanying lens model, calculate the magnifications and time delays, and infer physical properties for the source. We find the rest-frame UV continuum and emission lines to be dominated by the AGN, and the optical emission to be dominated by the host galaxy of modest stellar mass $M_{\star}\simeq10^{9.2}\,\mathrm{M}_{\odot}$. We also observe some variation in the AGN emission with time, which may suggest that the AGN used to be more active. This object adds a low-redshift counterpart to several relatively faint AGN recently uncovered at high redshifts with HST and JWST.
\end{abstract}

\begin{keywords}
quasars -- gravitational lensing: Strong -- galaxies: clusters: individual: MACS~J0035.4-2015 -- galaxies: nuclei -- cosmology: observations -- galaxies: Seyfert
\end{keywords}



\section{Introduction}
\label{sec:intro}
Active galactic nuclei (AGN) are galaxies hosting a supermassive black hole (SMBH) that is actively accreting matter at their center. The accretion process, in which potential and kinetic energy are transformed into thermal energy, results in very high luminosities \citep[for reviews see e.g.][]{Peterson2009elu..book..138P,Netzer2013peag.book.....N}.

A prominent sub-type of AGN are quasars, or \textit{quasi-stellar objects}, in which the emission is dominated by the accretion disk such that the object appears as a bright point source, with typical bolometric luminosities of $L_{\mathrm{acc}}\sim10^{44}-10^{48}\,\frac{\mathrm{erg}}{\mathrm{s}}$ \citep[see, e.g.,][]{Shen2020QuasarLF}. While many quasars and AGN are known \citep[e.g.][]{Banados2016100PS1Quasars,Lyke2020,Flesch2021}, only a relatively small fraction -- perhaps a few dozen -- are known to be multiply imaged \citep[at least with separations large enough to be seen with HST; e.g.,][]{Suyu2017MNRAS.468.2590S,Millon2020A&A...640A.105MCosmograil}, and only about a handful are known to be multiply imaged by galaxy clusters \citep[][]{Inada2003Natur.426..810I,Inada2006,Oguri2013MNRAS.429..482O,Dahle2013,sharon2017,Shu2018,Acebron2022QSO,Acebron2022QSO2,Martinez2022,Bogdan2022,Furtak2022AGN,Napier2023}.

Multiply imaged quasars play a significant role in astronomy. For example, they provide invaluable insight into the composition of the lens, which is dominated by a dark matter (DM) component of unknown nature. This insight is usually gained from flux anomalies between the different images of the quasar. Using the chromaticity due to microlensing, lensed quasars also enable insight into the source structure, such as the accretion disk and broad- and narrow-line regions \citep[e.g.,][]{Fian2021A&A...653A.109F,Mediavilla2011ApJ...730...16M,Rojas2014ApJ...797...61R}. In addition, thanks to lensing magnification, we are able to probe fainter quasars than the typical bright population, including objects that without lensing might not immediately be classified as potentially AGN-dominated based on their appearance. This is because at high redshifts faint galaxies can also appear as point-sources \citep[e.g.][]{Bouwens2017}. For a \textit{lensed} point-source, however, the size constraint is much stronger than for a blank-field point-source, since it translates into an even smaller size (typically several tens of pc) in the source plane \citep[e.g.][]{Furtak2022AGN}. Finally, perhaps the most notable role lensed quasars have played in the past decade is the measurement of the Hubble constant \citep[e.g.][]{Suyu2013measured,Courbin2018A&A...609A..71C,Wang2018z7Quasar,Wang_2021,Napier2023}, adding key constraints to the renewed tension in its local value compared to results from the Cosmic Microwave Background (CMB) \citep{Planck15,Riess2021ApJ...908L...6R,Wang_2021}. Because the path to each multiple image is different, the light arrival time for each image of a lensed quasar is different. That difference is in turn inversely proportional to the Hubble constant $H_{0}$ -- the expansion rate of the Universe.

Lensed quasars also constitute a rare example of X-ray sources multiply imaged by galaxy clusters, only few of which are known. As another example, several years ago \citet{Bayliss2020NatAs...4..159B} reported a multiply imaged X-ray source lensed by the galaxy cluster SPT-CLJ2344-4243, which in optical imaging appeared as a typical, elongated strongly lensed arc. Indeed, \citet{Bayliss2020NatAs...4..159B} concluded, based on various properties such as its morphology and emission line flux ratios, that in this case the X-ray emission comes from a star-formation region and is atypical for an AGN. This galaxy, however, is quite unique: out of the few X-ray sources known to be strongly lensed by clusters, most are indeed AGN --  and in particular, quasars.

Here, we report the discovery of a triply-imaged compact object detected in the \textit{Hubble Space Telescope} (HST) imaging of the galaxy cluster MACS~J0035.4-2015 \citep[hereafter MACS0035; $z_{\mathrm{d}}=0.352$;][]{Ebeling2010FinalMACS} taken several years ago for the \textit{REionization LensIng Cluster Survey} \citep[RELICS;][]{Coe2019RELICS}. The lensed object shows a nucleus of a prominent point-like morphology, is relatively bright for its size, and shows broad rest-frame ultra-violet (UV) emission features typical of an AGN. It has indeed previously been spectroscopically classified as a quasar in \citet[][]{Mainali2019PhDT.......198M}\footnote{PhD Thesis} but not analysed further. In addition, we detect this object in both X-ray and millimeter observations, which further confirms that it is an AGN. In this work we analyse the available data of this triply-imaged AGN and present its physical properties, adding to some recent samples of X-ray detected AGN in lensing cluster fields \citep{Bogdan2022,Uematsu2023arXiv230109275U}.

This paper is organised as follows: In \S \ref{sec:data} we describe the data used in this work and the lens model constructed for MACS0035. The source and its properties are presented and discussed in \S \ref{sec:results}, and the work is concluded in \S \ref{sec:conclusion}. Throughout this work, we use a standard flat $\Lambda$CDM cosmology with $H_0=70\,\frac{\mathrm{km}}{\mathrm{s}\,\mathrm{Mpc}}$, $\Omega_{\Lambda}=0.7$, and $\Omega_\mathrm{m}=0.3$. All magnitudes quoted are in the AB system \citep{Oke1983ABandStandards} and all uncertainties represent $1\sigma$ ranges unless stated otherwise. 

\begin{figure*}
    \centering
    \includegraphics[width=\textwidth]{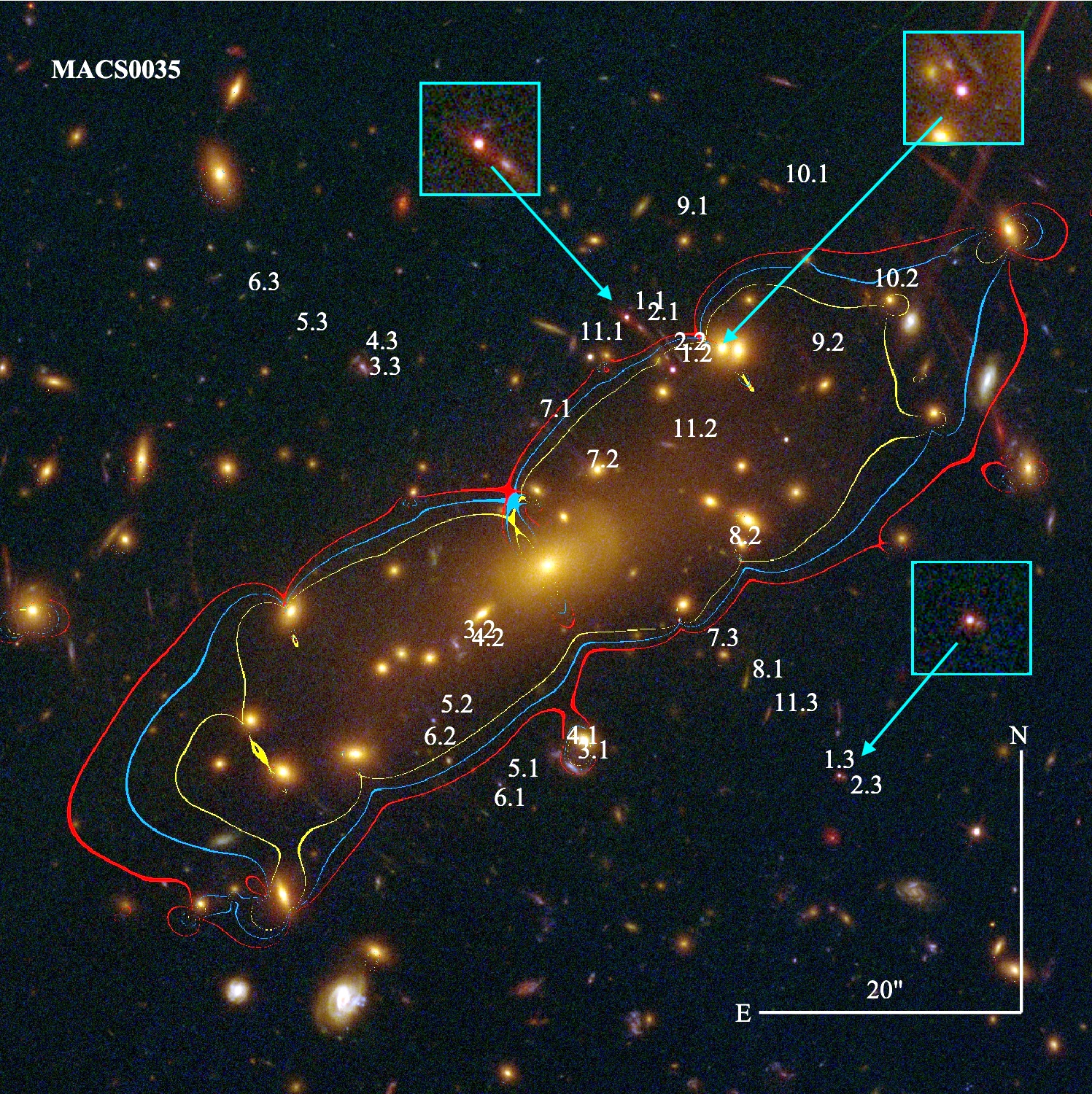}
    \caption{Color-composite image of MACS0035 constructed from the RELICS HST imaging of the cluster (Red: F105W+F125W+F140W+F160W; Green: F606W+F814W; Blue: F435W), centered on $\alpha=$~00:35:24, $\delta=$~-20:16:15. Strong lensing multiple images are numbered and labeled in white. The yellow, blue and red lines represent, respectively, the critical curves for source redshifts $z_{\mathrm{s}}=1.28$ (systems 3 \& 4), $z_{\mathrm{s}}=2.06$ (system 1; the multiply imaged AGN) and $z_{\mathrm{s}}\sim4.5$ (e.g. system 10), as computed from our SL model of the cluster (section~\ref{sec:SL}). Highlighted in cyan squares are the three multiple images of the AGN reported in this work.}
    \label{fig:cc}
\end{figure*}

\section{Data}\label{sec:data}
With the goal of finding high-redshift galaxies, the RELICS program \citep{Coe2019RELICS} imaged 41 massive galaxy clusters with HST to depths of about 26.5\,magnitudes (at $5\sigma$ in the F160W-band) in seven filters from the optical, with the \textit{Advanced Camera for Surveys} (ACS), to the near-infrared (NIR), with the \textit{Wide-field Camera Three} (WFC3). The RELICS observations of MACS0035 took place in 2016 November 10, and 2016 December 27. The final RELICS products that we use also include archival ACS data taken on 2005 September 13 in the F606W-band (PI: H.~Ebeling, Program~ID: 10491) and on 2013 June 2, in the F814W band (PI: H.~Ebeling, Program~ID: 12884). We use here the public data products from the RELICS program\footnote{\url{https://archive.stsci.edu/hlsp/relics}} which include, besides the drizzled broad-band images of MACS0035, a source catalog obtained with \texttt{SExtractor} \citep{BertinArnouts1996Sextractor} that contains photometry and photometric redshifts computed with \texttt{BPZ} \citep{Benitez2004}, and colour images made using \texttt{Trilogy} \citep{Coe2012A2261}. The details of the observations, data reduction and source extraction can be found in \citet{Coe2019RELICS}. Additional HST observations for MACS0035 were recently obtained on 2022 October 4 in the WFC3/UVIS F606W- and WFC3/IR F105W-bands (PI: P.~Kelly, Program~ID: 16729), which we use here as well for comparison.

In addition, we use ground-based spectroscopic data obtained with the \textit{Multi-Unit Spectroscopic Explorer} \citep[MUSE;][]{Bacon2010MUSE} on ESO's \textit{Very Large Telescope} (VLT), which are publicly available in the ESO Science Archive (Program~ID: 0103.A-0777; PI: A.~Edge; observation date: 2019-09-07 and 2019-07-02; exposure time: 2910\,s per observation), and with the \textit{Low Dispersion Survey Spectrograph 3} (LDSS3) on the \textit{Magellan Clay Telescope} as part of a RELICS spectroscopic follow-up program \citep[PI: K.~Sharon; observation date: 2017-07-27; exposure time: 4.8\,ks][]{Mainali2019PhDT.......198M}. MACS0035 is also part of the \textit{ALMA Lensing Cluster Survey} \citep[ALCS;][]{Kohno2019asrc.confE..64K,Fujimoto2023} which took high-resolution 1.2\,mm imaging ($1.42\arcsec\times1.03\arcsec$ beam) of 33 strong lensing clusters with the \textit{Atacama Large Millimeter/sub-millimeter Array} (ALMA). These data are also used in this work and are publicly available on the ESO Science Archive (Program~ID: 2018.1.00035.L; PI: K.~Kohno; observation date: 2019-03-12). Finally, we also use the existing X-ray imaging of MACS0035 taken with the \textit{Advanced CCD Imaging Spectrometer} (ACIS) aboard the \textit{Chandra X-ray Observatory} which is publicly available in the \textit{Chandra} Data Archive (Obs.~ID: 3262; PI: H.~Ebeling; observation date: 2003-01-22; exposure time: 21\,ks). 

\subsection{Gravitational lensing} \label{sec:SL}
We construct a lens model for the galaxy cluster MACS0035 using an updated version of the parametric lens modeling code from \citet{Zitrin2014CLASH25} that was presented in \citet{Pascale2022SMACS0723} and \citet{Furtak2022UNCOVER}. The model is based on 11 sets of multiple images identified in the HST imaging, comprising a total of 30 images (see Fig.~\ref{fig:cc}). In the MUSE observations of the field (see section~\ref{sec:data}), we detect a prominent double emission line feature -- consistent with the [\ion{O}{ii}]$\lambda\lambda3726,3729$\AA-doublet at $z=1.279\pm0.001$ -- for the two adjacent systems 3 and 4 (see Fig.~\ref{fig:cc}). The spectra are extracted and the redshifts measured with the same method that was used in \citet{Golubchik2022}. We note also that this redshift is independently confirmed by the \textit{Magellan} LDSS3 observations in \citet{Mainali2019PhDT.......198M}. For the other multiple image systems, which were not identified individually in the MUSE data, we use the photometric redshifts from the RELICS catalog as priors and allow their redshift to be freely optimised in the minimization. The cluster galaxies are modeled each as a dual pseudo isothermal elliptical mass distribution (dPIE; \citealt{Keeton2001models,Eliasdottir2007arXiv0710.5636E}) following typical luminosity scaling relations \citep{Jullo2007Lenstool} which are free to vary. One large cluster DM halo is used, modeled as a pseudo-isothermal elliptical mass distribution \citep[PIEMD; e.g.][]{Keeton2001models}. The model parameters are optimised via a long Monte-Carlo Markov Chain (MCMC) of several tens of thousand steps and with an input positional uncertainty of 0.5\arcsec\ for each multiple image. The best-fit model reproduces the position of the multiple images with a lens plane RMS of 0.7\arcsec. The images of the AGN studied in this work, i.e. system 1, in particular are reproduced with an RMS of 0.4\arcsec. The critical curves from the model for various redshifts are shown in Fig.~\ref{fig:cc}. We compute magnifications and time delays for each image of the AGN studied here and report them in Tab.~\ref{tab:images}. The best-fitting lensing redshift of this source is $z_{\mathrm{geo}}=2.1_{-0.2}^{+0.3}$, in excellent agreement with its spectroscopic redshift (see section~\ref{sec:spectroscopy}).

\subsection{Photometry} \label{sec:photometry}

\begin{figure*}
    \centering
    \includegraphics[width=\textwidth]{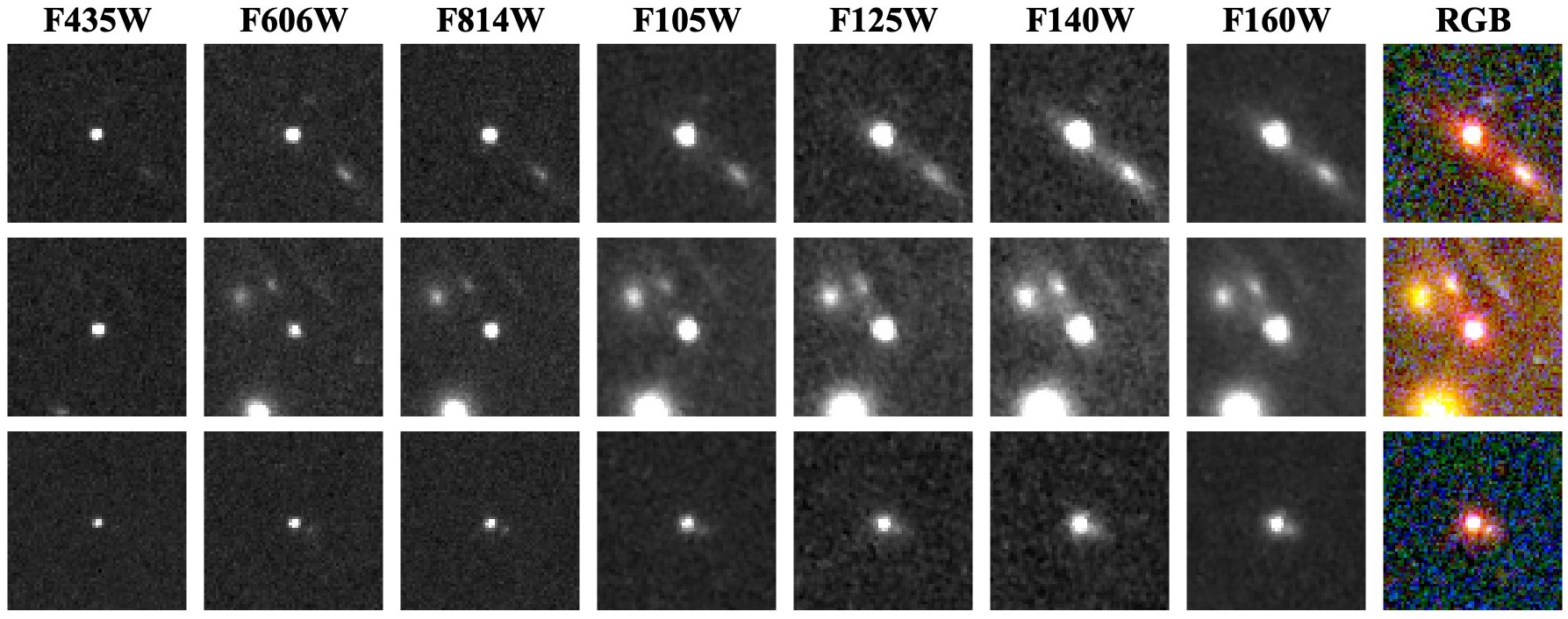}
    \caption{Image cutouts ($3.6\arcsec\times3.6\arcsec$) in each band of the three multiple images of the AGN. Images~1.1, 1.2 and~1.3 are shown from top to bottom in that order. The last column shows an enhanced composite-colour cutout from the same colour image as shown in Fig.~\ref{fig:cc}.}
    \label{fig:stamps}
\end{figure*}

\begin{table*}
\caption{Photometric and spectroscopic measurements of the three multiple images of the lensed AGN.}
\setlength{\tabcolsep}{2pt}
\begin{tabular}{lccccccccc}
\hline
\hline
&&&&&&&&&\\
ID & R.A. & Dec. & \textit{Chandra}/ACIS-I & F435W & F606W &F814W  &F105W & F125W & F140W \\
\hline
1.1   &   $8.8571211$ &   $-20.2570305$  &   $3.4_{-0.5}^{+7.0}\times10^{-15}$   &  $1.72\pm0.02$  &    $1.45\pm0.01$  &    $2.56\pm0.02$  &    $3.38\pm0.02$  &    $3.80\pm0.02$  &    $4.71\pm0.02$  \\
1.2   &   $8.8560826$ &   $-20.2581397$  &   $9.2_{-5.2}^{+14.1}\times10^{-15}$  &  $2.52\pm0.02$  &    $0.60\pm0.02$  &    $2.18\pm0.07$  &    $3.13\pm0.10$  &    $3.64\pm0.11$  &    $4.25\pm0.14$  \\
1.3   &   $8.8523185$ &   $-20.2667358$  &   $6.3_{-2.9}^{+10.6}\times10^{-15}$ &  $0.50\pm0.01$  &    $0.43\pm0.01$  &    $0.76\pm0.01$  &    $0.85\pm0.01$  &    $1.27\pm0.02$  &    $1.67\pm0.02$  \\
\hline
\hline
&&&&&&&&&\\
~~~F160W & ALMA~Band~6 & $z_{\mathrm{phot}}$~[95\% C.I.] & $\mu$ & $\Delta t$ [d] & $\mathrm{EW}_{0,\mathrm{CIV}\lambda1550\mathring{\mathrm{A}}}$ & $\mathrm{EW}_{0,\mathrm{HeII}\lambda1640\mathring{\mathrm{A}}}$ & $\mathrm{EW}_{0,\mathrm{CIII]}\lambda1909\mathring{\mathrm{A}}}$&\\
\hline
$5.50\pm0.02$  &    $190\pm55$  &   $1.84~[1.81,2.01]$  &   $5.2_{-0.7}^{+1.0}$   &   -                      &   $45.6\pm0.3$\,\AA   &   $3.7\pm0.4$\,\AA    &   $27.5\pm0.5$&\\
$4.55\pm0.15$  &    $221\pm55$  &   $1.14~[1.10,1.23]$  &   $6.1_{-0.6}^{+1.5}$   &   $167_{-64}^{+9}$       &   -   &   -   &   -&\\
$1.92\pm0.01$  &    $231\pm61$  &   $2.12~[2.04,2.24]$  &   $3.6_{-0.6}^{+1.5}$   &   $-8815_{-198}^{+1222}$ &   -   &   -   &   -&\\
\hline
\hline
\end{tabular}
\par\smallskip
\begin{justify}
    \texttt{Note.} -- Optical, NIR and millimeter fluxes, given in $\mu$Jy, are measured in the RELICS HST/ACS+WFC3 images and the ALMA data (see section~\ref{sec:data}). The \textit{Chandra} X-ray fluxes are integrated fluxes from $0.5-7.0$\,keV in units of $\frac{\mathrm{erg}}{\mathrm{s}\,\mathrm{cm}^2}$ with their 90\% errors. All fluxes are observed, i.e. not de-magnified yet. The photometric redshift estimates are from the RELICS catalog \citep[][note that the first two images are contaminated by near-by galaxies and ICL in the RELICS catalog which also affects the photometric redshift estimates]{Coe2019RELICS}. The magnifications $\mu$ and time delays $\Delta t$ are computed with our lens model described in section~\ref{sec:SL}. The latter are in days relative to the first image, 1.1. The last three columns show the rest-frame EWs of the emission lines measured in the LDSS3 spectrum in section~\ref{sec:spectroscopy}. Note that we do not resolve the \ion{C}{iv} and \ion{C}{iii}] doublets.
\end{justify}
\label{tab:images}
\end{table*}

\begin{figure*}
    \centering
    \includegraphics[width=0.3\textwidth]{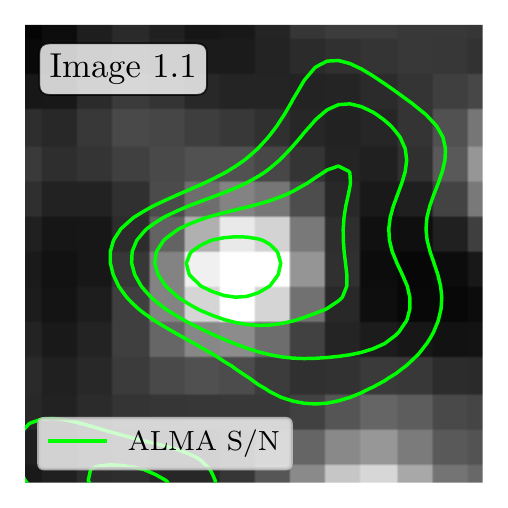}
    \includegraphics[width=0.3\textwidth]{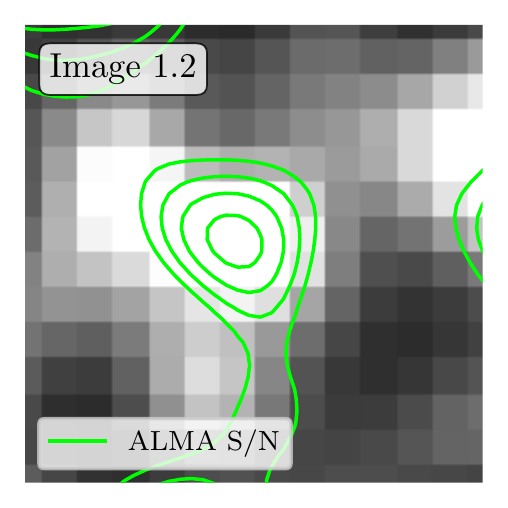}
    \includegraphics[width=0.3\textwidth]{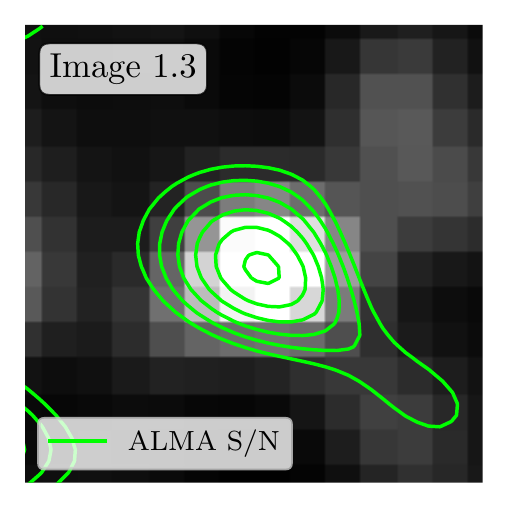}
    \caption{Cutouts ($7\arcsec\times7\arcsec$; 0.5\arcsec/pix) of the three images of the lensed AGN in the \textit{Chandra}/ACIS-I X-ray data ($0.5-7$\,keV; smoothed with a 3-pixel Gaussian kernel) overlaid with the ALMA Band~6 (1.2\,mm) signal-to-noise contours in green (ranging from 0.5 to 3 in steps of 0.5). All three images are clearly detected in both X-ray and millimeter wavelengths.}
    \label{fig:X-ray}
\end{figure*}

While all three images of the AGN (see Figs.~\ref{fig:cc} and~\ref{fig:stamps}) are detected in the RELICS catalog (section~\ref{sec:data}), we note upon examination of the corresponding segmentation map that the detections are blended with another nearby multiple image system (Fig.~\ref{fig:cc}) in the catalog. Because of that, we measure the photometry of the three images of the AGN with \texttt{photutils} \citep[\texttt{v1.6.0};][]{photutils22} in circular apertures of 0.5\arcsec\ diameter and a local background annulus, both of which are carefully chosen to avoid contamination by the close-by companion (but does include the thin red envelope around the nucleus; see Figs.~\ref{fig:cc} and~\ref{fig:stamps}). The aperture fluxes are then corrected using aperture correction factors that take the point-spread-function (PSF) in each band into account. The resulting fluxes in each band are listed for each image in Tab.~\ref{tab:images} and cutouts of the images in the individual bands can be seen in Fig.~\ref{fig:stamps}. Despite some variation, possibly attributed to contribution from nearby galaxies especially for image~1.2 (although see more discussion about variability in section \ref{sec:results}), the flux ratios of the images seem to be broadly consistent over all bands and concur with the magnification ratios within the $1\sigma$-uncertainties.

In the ALMA continuum map, we identify a $\sim4\sigma$ source within the ALMA beam size ($\sim1\arcsec$) around each multiple image position. We thus attribute these ALMA sources to be the rest-frame far-infrared (FIR) counterparts of these multiple images and measure the 1.2\,mm flux densities from their peak pixel counts, as reported in Tab.~\ref{tab:images}. Note that image~1.2 was previously detected as an ALMA source in a joint ALCS and \textit{Herschel Space Observatory} \citep[][]{Pilbratt2010} study by \citet{Sun2022} but did not show any \textit{Herschel} flux above the detection threshold. All three images of the lensed AGN are also detected in \textit{Chandra} X-ray $0.5-7$\,keV maps of the field (see section~\ref{sec:data}) although with relatively low count statistics ($<10$ net photons). The integrated X-ray fluxes in the $0.5-7$\,keV energy band, corrected for galactic absorption, are also given in Tab.~\ref{tab:images} and we show the X-ray map overlaid with the ALMA contours of the object in Fig.~\ref{fig:X-ray}. The X-ray fluxes are measured in 2\arcsec\ apertures with local background estimates which were corrected for the \textit{Chandra} PSF and off-axis angle. The ALMA and \textit{Chandra} flux ratios between multiple images do not concur with the magnification ratios, nor with the optical flux ratios, i.e. the least magnified image~1.3 is the brightest rather than the most magnified one, image~1.2. This may indicate variable AGN activity, as we discuss further in section~\ref{sec:results} but note also that the uncertainties on these measurements are very large.

Finally, the RELICS catalog also contains \texttt{BPZ} photometric redshifts that are also listed in Tab.~\ref{tab:images} for completeness.

\subsection{Spectroscopy} \label{sec:spectroscopy}

\begin{figure}
    \centering
    \includegraphics[width=\columnwidth]{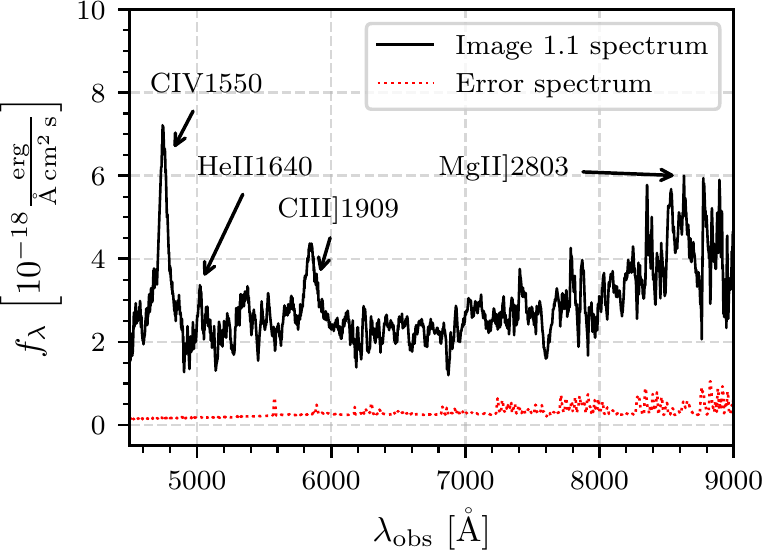}
    \caption{LDSS3 spectrum of image~1.1 (black) and its $1\sigma$ uncertainty (red) which was published in \citet{Mainali2019PhDT.......198M}. The two prominent broad emission lines are identified as the \ion{C}{iv}$\lambda1550$\AA\ and \ion{C}{iii}]$\lambda1909$\AA\ (unresolved) doublets at $z_{\mathrm{spec}}=2.063\pm0.005$. There is also a \ion{He}{ii}$\lambda1640$\AA\ detection and a hint at the \ion{Mg}{ii}]$\lambda2803$\AA\ doublet towards the high-wavelength end.}
    \label{fig:spectrum}
\end{figure}

The LDSS3 mask (see section~\ref{sec:data}) targeted one of the three known images of the source, image~1.1, among other objects of interest in the field \citep[see][]{Mainali2019PhDT.......198M}. These data were reduced and presented in \citet{Mainali2019PhDT.......198M}, who reported the detection of several broad UV emission lines which enabled the identification of this object as a quasar at redshift $z_{\mathrm{spec}}\simeq2.069$.

As can be seen in Fig.~\ref{fig:spectrum}, the spectrum features several prominent emission lines. The two brightest ones are the \ion{C}{iv}$\lambda1550$\AA\ and \ion{C}{iii}]$\lambda1909$\AA\ (unresolved) doublets at $z\simeq2.1$, and there is also a somewhat weaker \ion{He}{ii}$\lambda1640$\AA\ emission. We use the \texttt{specutils} package \citep[\texttt{v1.9.1};][]{specutils22} to compute the equivalent widths (EWs) and perform a Gaussian fit to the detected lines in order to measure line centers and full-width-half-maxima (FWHM). We derive a spectroscopic redshift of $z_{\mathrm{spec}}=2.063\pm0.005$ for this object, consistent with \citet{Mainali2019PhDT.......198M}. Given their high ionization potential and broad (rest-frame) line widths, i.e. FWHM of $4696\pm78\,\frac{\mathrm{km}}{\mathrm{s}}$, $5108\pm191\,\frac{\mathrm{km}}{\mathrm{s}}$, and $1337\pm263\,\frac{\mathrm{km}}{\mathrm{s}}$ respectively, these lines indeed confirm this object as an AGN. Note that since we do not resolve the \ion{C}{iv} and \ion{C}{iii}] doublets, the quoted line widths nominally represent upper limits, but the true values should not be significantly smaller given that the line widths (tens of \AA) are much larger than the separation between the two doublet components ($\simeq2$\,\AA). The measured EWs are listed in Tab.~\ref{tab:images}. We also find a tentative \ion{Mg}{ii}]$\lambda2803$\AA\ detection at the same redshift, but it is too uncertain to derive a robust line center and EW (it is also possible that the higher continuum seen towards redder wavelengths is in part due to complex \ion{Fe}{ii} emission; see e.g. \citealt{Sameshima2011}). Note that the \ion{C}{iv}$\lambda1550$\AA\ and \ion{C}{iii}]$\lambda1909$\AA\ lines can also be seen in the MUSE spectra, though at very low signal-to-noise ($\sim1.5$; when combining the signal from the different multiple images), and corroborate the redshift and the large line width measurements from the LDSS3 data.
 
\section{A lensed AGN -- Physical properties of the source} \label{sec:results}

\begin{figure*}
    \centering
    \includegraphics[width=0.49\textwidth]{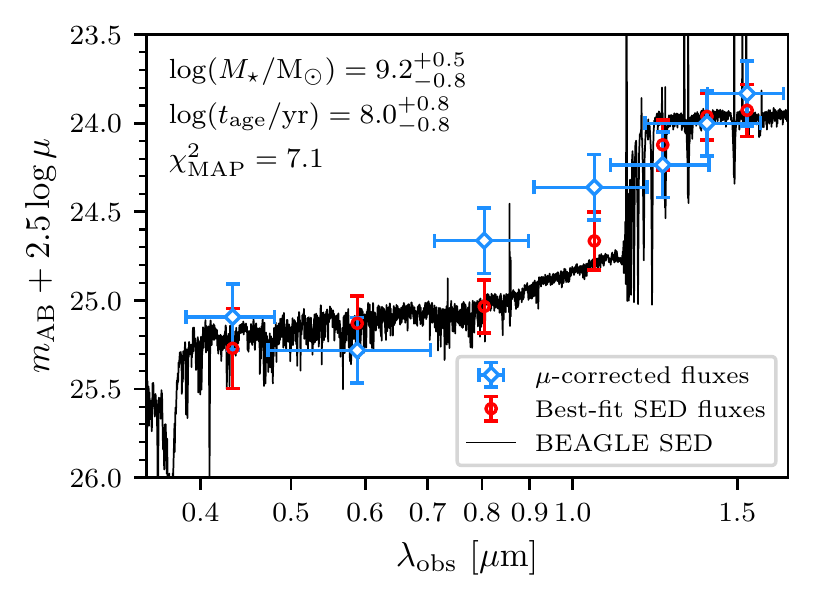}
    \includegraphics[width=0.49\textwidth]{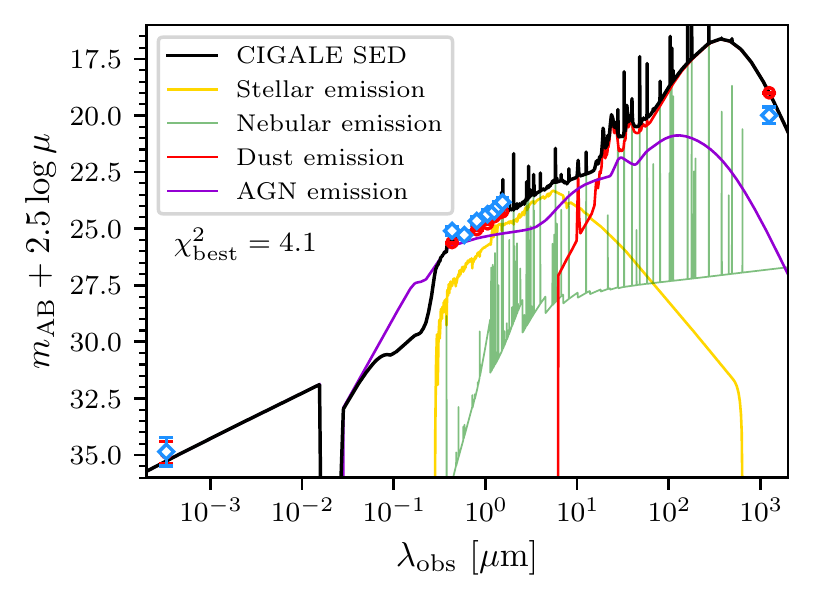}
    \caption{SED-fitting results with \texttt{BEAGLE-AGN} (\textit{left}) and \texttt{CIGALE} (\textit{right}). The maximum-a-posteriori (MAP) SEDs are shown in black, the de-magnified photometry of image~1.1 is shown in blue, and the best-fitting model fluxes in each band are shown in red. The \texttt{BEAGLE-AGN} SED-fit is performed with a hybrid combination of a star-forming galaxy with a type~II AGN component, similar to \citet{Endsley2022AGNcandidate_MNRAS.512.4248E} and \citet{Furtak2022AGN}. Our \texttt{CIGALE}-fit combines galaxy and type~I AGN templates and in addition accounts for X-ray and FIR (i.e. dust) emission.}
    \label{fig:SED}
\end{figure*}

\begin{figure}
    \centering
    \includegraphics[width=\columnwidth]{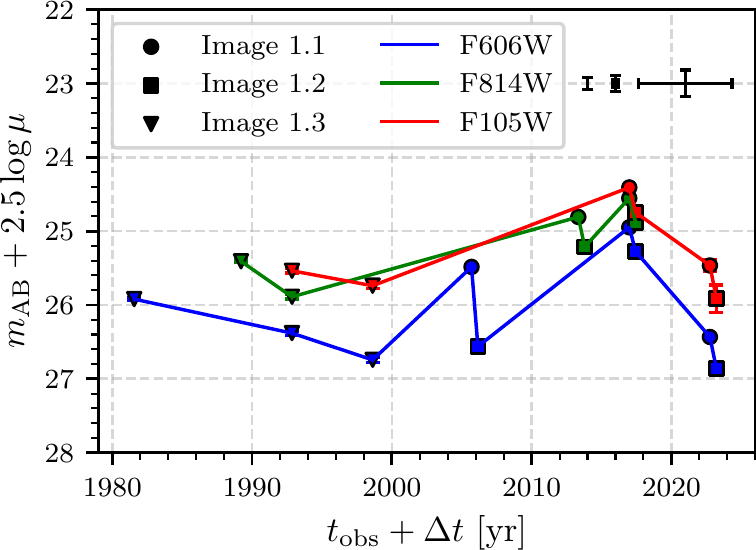}
    \caption{Rest-frame UV variation of the AGN. We utilise the different visits (see section~\ref{sec:data}) and image multiplicity to gain insight into the variability of the AGN. Using the lens model's magnifications $\mu$ and time delays $\Delta t$ listed in Tab.~\ref{tab:images}, we de-magnify and time-shift each image in each visit to obtain a measurement of the source in time. We repeat this for the three bands in which largely separated visits took place, i.e. F606W (blue), F814W (green), F105W (red). All measurements of a certain band are connected by a line to help track the variability. Note that the last F606W measurement is from WFC3/UVIS and not ACS. Measurements derived from image 1.1 are marked as circles, measurements derived from image 1.2 are marked as squares, and measurements derived from image 1.3 are marked as triangles. The error-crosses in the upper-right corner show the typical systematic offsets resulting from the magnification and time-delay uncertainties for each image from left to right: Image 1.1, 1.2, and 1.3 respectively). Note these quantities however should be roughly fixed for all epochs, and do not affect the variability. The timeline is set such that image~1.1 is at zero time-shift with respect to the observation dates as listed in Tab.~\ref{tab:images}.} 
    \label{fig:variability}
\end{figure}

In the colour image shown in Fig.~\ref{fig:cc}, several different features are noticeable in the multiple images of the AGN, especially in the two highly magnified ones (1.1 and 1.2; see Tab.~\ref{tab:images}). The first most notable feature is the blue/white, point-like nucleus. Given it is a point-source in the blue optical bands, verified by dedicated \texttt{GALFIT} \citep{Peng2010AJ....139.2097P} analysis, we can place an approximate upper limit on its size by adopting a nominal 0.1\arcsec\ FWHM PSF in the HST imaging. Taking the magnification and distance into account, this translates into a size of $\lesssim200$\,pc in the source plane, or $r_{e}\lesssim100$\,pc. The second feature is identified as a thin redder envelope around the blue nucleus, seen in the longer-wavelength HST/WFC3 bands, which is only marginally point-like, i.e. it presents a slight stretch in the direction of the arc (the direction of the shear). This means that the red emission likely comes from a slightly larger region than the nucleus, which, following similar arguments as above and a \texttt{GALFIT} measurement in the F160W where this feature is more pronounced, should not be larger than about $\lesssim180$\,pc in radius. The measured surface brightness of this emission is $24.12\pm0.05\,\mathrm{mag}/\mathrm{arcsec}^2$ in the F160W-band ($\sim5000$\,\AA\ in the galaxy's rest-frame at $z\simeq2.06$) and it looks to be of a spherical morphology in the least-sheared image (image~1.3). The last feature is a relatively blue and bright ($22.05\pm0.01$\,AB in the F160W-band) patch $\sim1.2\arcsec$ from the nucleus along the arc which corresponds to a de-magnified projected distance of about 3.3\,kpc (image system 2 in Fig.~\ref{fig:cc}). Assuming this clump has the same redshift as the nucleus, it could be an outer part of the host, probably a star-forming knot, or perhaps more likely, a small interacting companion.

At the redshift of the source, the 1450\,\AA\ emission is contained in the F435W-band. We therefore use the photometry of image~1.1 measured in that filter to derive a de-magnified UV luminosity of $M_{\mathrm{UV},1450}=-19.7\pm0.2$. This is relatively faint for a quasar-like object, and could potentially fall under the category of Seyfert galaxies, also given part of the host is seen. We refer to the object as a quasar-like throughout in the simplest sense that it is an AGN with an apparent quasi-stellar appearance. By assuming a single modified black-body with a typical dust temperature of $T_{\rm d}=47$\,K and dust-emissivity spectral index $\beta_{\rm d}=1.6$ \citep{Beelen2006ApJ...642..694B} scaled to the ALMA detection (see Tab.~\ref{tab:images}), we derive an FIR luminosity of $L_{\mathrm{FIR}}=(2.3\pm0.6)\times10^{11}\,\mathrm{L}_{\odot}$. Applying the \citet{Rieke2009} relation, this results in a dust-obscured star-formation rate (SFR) of $\psi_{\mathrm{FIR}}=10.5\pm6.5\,\frac{\mathrm{M}_{\odot}}{\mathrm{yr}}$ which agrees well with the optical emission line inferred SFR of the host derived below. The X-ray luminosity of image~1.1 is $L_\mathrm{X}=2.1_{-0.5}^{+4.4}\times10^{43}\,\frac{\mathrm{erg}}{\mathrm{s}}$. Using the relation between the UV luminosity, the \ion{C}{iv}$\lambda1550$\AA\ FWHM, the line ratio between the \ion{C}{iv}$\lambda1550$\AA\ and \ion{C}{iii}]$\lambda1909$\AA\ lines and the black hole mass from \citet{PopovicCIVmass} based on \citet{MejiaRestrepo2016AGNmass} and \citet{Ge2019AGNCIV}, we obtain an estimate of the black hole mass of $\log(M_{\mathrm{BH}}/\mathrm{M}_{\odot})=7.22\pm0.05$. Together with the stellar mass of the host galaxy found below, this broadly agrees with the typical $M_{\mathrm{BH}}-\sigma_{\star}$ (i.e. black hole mass-velocity dispersion of stars in the galaxy) or black hole mass-luminosity relation established for local AGN \citep[e.g.][]{Gultekin2009}. This relation has not yet been probed out to $z\gtrsim1$ and in particular not at these masses \citep[e.g.][]{Canalizo2012,Park2015,Schindler2016S,Li2021}. Note that since the spectrum has only been measured for image~1.1 (see section~\ref{sec:spectroscopy}), we cannot at this stage derive the black hole mass for the other two images.

To constrain the physical properties of this object, we perform a spectral energy distribution (SED) analysis by fitting the photometry of image~1.1 (see Tab.~\ref{tab:images}) after correcting it for magnification, and fix the fit to the spectroscopic redshift (see section~\ref{sec:spectroscopy}), with two distinct codes: We use the \texttt{BayEsian Analysis of GaLaxy sEds} (\texttt{BEAGLE}) tool \citep[][]{chevallard16} on the broad-band HST data and, in order to also fold-in the \textit{Chandra} and ALMA detections, the \texttt{Code Investigating GAlaxy Emission} code \citep[\texttt{CIGALE};][]{Boquien2019,Yang2020CIGALE,Yang2022CIGALE}. Since the flux ratios of the three images concur with the magnification ratios as described in section~\ref{sec:photometry}, the de-magnified photometry is identical within the uncertainties for each image and would thus yield the same SED-fit. For \texttt{BEAGLE}, following the approach of \citet{Endsley2022AGNcandidate_MNRAS.512.4248E} and \citet{Furtak2022AGN}, we use hybrid templates consisting of a galaxy component and an AGN component \citep{feltre2016} newly developed for \texttt{BEAGLE} \citep[this update is called \texttt{BEAGLE-AGN};][currently only a type~II AGN template is available in this library]{vidal-garcia22}. We assume a delayed star-formation history (SFH), a \citet{Charlot2000} dust attenuation law and the \citet{inoue14} intergalactic-medium (IGM) absorption models in the \texttt{BEAGLE-AGN} fit. In \texttt{CIGALE}, which also includes the possibility to fit a type~I AGN, we assume a delayed SFH and a Calzetti dust attenuation law \citep{Calzetti2000} for the host galaxy, the \texttt{Skirtor2016} \citep{Stalevski2012,Stalevski2016} AGN emission models for the AGN component and the \citet{Dale2014} dust emission templates. Note that we fix the \texttt{CIGALE} templates to a type~I AGN. The resulting best-fit SEDs are shown in Fig.~\ref{fig:SED}.

The \texttt{BEAGLE-AGN}-fit yields a moderate stellar mass and age of $\log(M_{\star}/\mathrm{M}_{\odot})=9.2_{-0.8}^{+0.5}$ and $t_{\mathrm{age}}=100_{-84}^{+630}$\,Myr, respectively, for the host galaxy. It additionally finds a significant current SFR (representing the last 10\,Myr) of $\psi_{10\,\mathrm{Myr}}\simeq13\,\frac{\mathrm{M}_{\odot}}{\mathrm{yr}}$ and a dust attenuation of $A_V\simeq0.9$. Note that this SFR estimates agrees well with the FIR luminosity inferred SFR value above. While this SED-fit with \texttt{BEAGLE-AGN} reproduces the broad-band photometry reasonably well within the uncertainties, it does not, however, reproduce the strong rest-frame UV emission lines measured in section~\ref{sec:spectroscopy} (see Tab.~\ref{tab:images}). This is not surprising, since the \texttt{BEAGLE-AGN} templates only model a type~II AGN whereas the broad emission lines clearly originate from a type~I AGN. The resulting best-fitting \texttt{CIGALE} host-galaxy parameters are $\log(M_{\star}/\mathrm{M}_{\odot})=9.50\pm0.26$ and $t_{\mathrm{age}}=1024\pm625$\,Myr which agree with the \texttt{BEAGLE-AGN} results within the uncertainties. The best-fitting \texttt{CIGALE} SED further suggests an accretion luminosity of $L_{\mathrm{acc}}\simeq7\times10^{43}\,\frac{\mathrm{erg}}{\mathrm{s}}$. As can be clearly seen in the right-hand panel of Fig.~\ref{fig:SED}, the rest-frame UV emission is dominated by the type~I AGN component ($\simeq72$\,\% of the integrated flux) whereas the rest-frame optical is dominated by the host galaxy ($\simeq41$\,\% AGN contribution). The latter is also consistent with the fact that the AGN contribution to the integrated flux inferred with \texttt{BEAGLE-AGN} is $30\pm15$\,\%.

The rest-frame optical HST flux ratios concur with the magnification ratios from the lens model. However, the X-ray, rest-frame UV, and FIR fluxes do not (see section~\ref{sec:photometry}). This concurs with the picture that the optical emission originates mostly from the host galaxy, and thus is stable, whereas the X-ray and UV (and potentially the FIR) originate directly from the AGN (which also contributes significantly to the heating of the dust) and are therefore sensitive to variations in its activity. The X-ray emission typically varies on different time-scales than the FIR emission due to the distance between the AGN and the dust around it (of order $\sim$several pc). Since image~1.3 is the first to arrive, by $\sim24$\,yr according to our SL model (see Tab.~\ref{tab:images} and section~\ref{sec:SL}), its enhanced X-ray and FIR emission may therefore suggest that the AGN was more active at that time. Note however that since uncertainties on both the X-ray and ALMA measurements, and the magnifications, are quite large (see Tab.~\ref{tab:images}), and since the X-ray and FIR data were taken almost 15 years apart, this conclusion remains uncertain. In addition, the X-ray emission typically varies on different time-scales than the FIR emission due to the distance between the AGN and the dust around it (of order $\sim$several pc). Regarding the possible AGN variability, we also make use of the fact that the cluster was observed in some of the blue HST filters in several epochs. For example, the imaging in the F606W-band was repeated three times over nearly two decades (two with ACS, and the most recent with WFC3/UVIS; see section~\ref{sec:data}). We do find signatures of variability also in these HST data when re-measuring the photometry of the three images in the single-epoch frames: The rest-frame UV measurements of the three AGN images vary by as much as $\sim1-1.5$\,magnitudes between the different epochs, as can be seen in Fig.~\ref{fig:variability}. Note that while AGN typically vary more in the rest-frame UV than in the optical, we do not have enough epochs in the rest-frame optical bands (i.e. F125W and beyond) to confirm this for the object studied here. That being said, the strongest variation is clearly seen in the bluest (F606W) band.

We thus conclude that the object studied here is likely a un-obscured moderate-luminosity ("Seyfert"-level) AGN, possibly undergoing a merger with a companion galaxy at Cosmic Noon ($z\simeq2$). It is possible that the enhanced AGN activity is associated with merger-driven AGN growth. This phenomenon is well documented locally \citep[e.g.][]{Comerford2015,Trakhtenbrot2017ApJ...836L...1T,Ricci2021}, but rarely with such compact hosts at such small projected distances. Although intrinsically rather faint and compact, the lensing magnification of this AGN could ultimately allow a privileged view of the dynamics of this system. This would allow us to better understand potential changes in AGN and galaxy growth over cosmic time, e.g. with future dedicated integral-field-unit (IFU) spectroscopic observations.

\section{Conclusion} \label{sec:conclusion}
We present a triply-imaged, relatively faint ($M_{\mathrm{UV},1450}=-19.7\pm0.2$) quasar-like object at $z_{\mathrm{spec}}=2.063\pm0.005$, lensed by the galaxy cluster MACS0035. In addition to its nucleus' point-like appearance, the source is detected in both X-ray and millimeter wavelengths, which, together with its broad ($\mathrm{FWHM}\sim1000-5000\,\frac{\mathrm{km}}{\mathrm{s}}$) high-ionization \ion{C}{iv}$\lambda1550$\AA, \ion{He}{ii}$\lambda1640$\AA\ and \ion{C}{iii}]$\lambda1909$\AA\ emission lines confirm it is an AGN, adding to less than a handful of known quasars multiply imaged by galaxy clusters. We run \texttt{BEAGLE-AGN} and \texttt{CIGALE} on the photometry of the source to derive its physical properties. We find that its host galaxy is relatively young, about $\sim100$\,Myr to $\sim1$\,Gyr old, with a moderate stellar mass of $\log(M_{\star}/\mathrm{M}_{\odot})=9.2_{-0.8}^{+0.5}$, and that about 40\,\% of the rest-frame optical emission originates from the AGN, which dominates however in the UV. We also note that while the HST bands largely agree with the magnification ratios from our lens model, both the X-ray and the millimeter observations seem to show a slight excess in the first-arrived image, suggesting it may have been caught in a particularly active phase of the AGN, even though the uncertainties are very large. We do find however substantial variability in the rest-frame UV emission of the AGN by using archival data from the past 20 years. The time delay between the two nearby images (1.1 and 1.2) is of order several months, which, given this variability, may thus make this system a useful target for constraining $H_{0}$. Future spectroscopic observations of this object could possibly further examine this and allow constraints on cosmological parameters thanks to the gravitational time delays. High-resolution IFU observations of this object might further yield insight into the dynamics of this potentially merging Seyfert galaxy system at $z\sim2$. 

The AGN we present here adds to three other lensed -- albeit not multiply imaged -- AGN recently detected in the ALCS survey \citep{Uematsu2023arXiv230109275U}. Moreover, this AGN adds to a few other AGN and intriguingly faint quasar candidates that were recently detected with the JWST, out to higher redshifts \citep[e.g.][]{Bogdan2022,Endsley2022arXiv220814999E,Onoue2022arXiv220907325O,Furtak2022AGN,Kocevski2023,Larson2023,Harikane2023}. Despite, or perhaps because of their observational rarity, faint quasars are particularly interesting. If indeed the quasar luminosity function remains steep towards the faint end \citep[e.g.][]{Glikman2011ApJ...728L..26G,Niida_2020}, this may indicate that quasars, or perhaps hybrid objects \citep[e.g.][]{Laporte2017,Mainali2017,Fujimoto2022Natur.604..261F}, contribute more heavily to the UV background \citep[][]{MadauHaardt2015} than is commonly assumed, and may have thus played a more significant role in the reionisation of the Universe. 

\section*{Acknowledgements}
The authors would like to thank the anonymous referee for their useful comments, which helped to improve the manuscript. A.Z. thanks Benny Trakhtenbrot for a useful discussion. L.J.F. and A.Z. acknowledge support by grant 2020750 from the United States-Israel Binational Science Foundation (BSF) and grant 2109066 from the United States National Science Foundation (NSF), and by the Ministry of Science \& Technology, Israel. J.C. acknowledges funding from the ``FirstGalaxies'' Advanced Grant from the European Research Council (ERC) under the European Union’s Horizon 2020 research and innovation program (Grant agreement No.~789056). E.C.L. acknowledges support of an STFC Webb Fellowship (ST/W001438/1). K.K. acknowledges the support by JSPS KAKENHI Grant Number JP17H06130 and the NAOJ ALMA Scientific Research Grant Number 2017-06B. D.E. acknowledges support from a Beatriz Galindo senior fellowship (BG20/00224) from the Spanish Ministry of Science and Innovation, projects PID2020-114414GB-100 and PID2020-113689GB-I00 financed by MCIN/AEI/10.13039/501100011033, project P20-00334  financed by the Junta de Andaluc\'{i}a, and project A-FQM-510-UGR20 of the FEDER/Junta de Andaluc\'{i}a-Consejer\'{i}a de Transformaci\'{o}n Econ\'{o}mica, Industria, Conocimiento y Universidades. G.E.M. acknowledges financial support from the Villum Young Investigator grant 37440 and 13160 and the The Cosmic Dawn Center (DAWN), funded by the Danish National Research Foundation under grant No. 140. F.E.B. acknowledges support from ANID-Chile BASAL CATA FB210003, FONDECYT Regular 1200495 and 1190818, and Millennium Science Initiative Program--ICN12\_009. K.K.K. acknowledges support from the Knut and Alice Wallenberg Foundation.

This work is based on observations made with the NASA/ESA \textit{Hubble Space Telescope} (HST). The data were obtained from the \texttt{Barbara A. Mikulski Archive for Space Telescopes} (\texttt{MAST}) at the \textit{Space Telescope Science Institute} (STScI), which is operated by the Association of Universities for Research in Astronomy (AURA) Inc., under NASA contract NAS~5-26555 for HST. This research has made use of data obtained from the \textit{Chandra} Data Archive and software provided by the \textit{Chandra} X-ray Center (CXC) in the application packages \texttt{CIAO} and \texttt{Sherpa}. This work is also based on observations made with ESO Telescopes at the La Silla Paranal Observatory and the \textit{Atacama Large Millimeter/sub-millimeter Array} (ALMA), obtained from the ESO Science Archive. ALMA is a partnership of ESO (representing its member states), NSF (USA) and NINS (Japan), together with NRC (Canada), MOST and ASIAA (Taiwan), and KASI (Republic of Korea), in cooperation with the Republic of Chile. The Joint ALMA Observatory is operated by ESO, AUI/NRAO and NAOJ. The National Radio Astronomy Observatory (NRAO) is a facility of the NSF operated under cooperative agreement by Associated Universities Inc. This paper is based on data gathered with the 6.5\,m \textit{Magellan} Telescopes located at Las Campanas Observatory, Chile, awarded through the University of Michigan.

This research made use of \texttt{Astropy},\footnote{\url{http://www.astropy.org}} a community-developed core \texttt{Python} package for Astronomy \citep{astropy13,astropy18} as well as the packages \texttt{NumPy} \citep{vanderwalt11}, \texttt{SciPy} \citep{virtanen20}, \texttt{spectral-cube} \citep{spectral-cube14} and the astronomy library for \texttt{MATLAB} \citep{maat14}. The \texttt{Matplotlib} package \citep{hunter07} was used to create some of the figures in this work.

\section*{Data Availability}
The HST data used in this work are publicly available ion the \texttt{MAST} archive, under program IDs~12884, 10491, 14096 and~16729, and on the RELICS website\footnote{\url{https://archive.stsci.edu/hlsp/relics}}. The \textit{Chandra} X-ray data are available on the \textit{Chandra} Data Archive\footnote{\url{https://cda.harvard.edu/chaser}} under observation ID~3262. The ESO/VLT and ALMA data can be obtained from the ESO Science Archive\footnote{\url{http://archive.eso.org/scienceportal/home}} under program IDs~103.A-0777 and~2018.1.00035.L. Finally, the \textit{Magellan} data and the lens models used in this work will be shared by the authors upon request.


\bibliographystyle{mnras}
\bibliography{MyBiblio} 



\appendix


\bsp	
\label{lastpage}
\end{document}